# Behaviors of Martian CO₂-driven dry climate system and conditions for atmospheric collapses

Yasuto Watanabe[1,†,*], Eiichi Tajika[1], Arihiro Kamada[2]

[1]Department of Earth and Planetary Science, Graduate School of Science, The University of Tokyo, Tokyo, Japan
[2]Department of Geophysics, Graduate School of Science, Tohoku University, Sendai, Japan
[†]Present affiliation: Meteorological Research Institute, Japan Meteorological Agency, Tsukuba, Japan
[*]Corresponding author: Yasuto Watanabe (yasuto.watanabe.wess@gmail.com). Department of Atmosphere, Ocean, and Earth System Modeling Research, Meteorological Research Institute, Japan Meteorological Agency. 1–1 Nagamine, Tsukuba, Ibaraki, Japan. Tel. +81-29-853-8648

## Abstract

The present Martian climate is characterized by a cold and dry environment with a thin atmosphere of carbon dioxides ($CO_2$). In such conditions, the planetary climate and habitability are determined by the distribution of $CO_2$ between exchangeable reservoirs, that is the atmosphere, ice caps, and regolith. This produces unique responses of the Martian $CO_2$-driven climate system to variations of astronomical forcings. Specifically, it has been shown that the phenomenon called an atmospheric collapse occurs when the axial obliquity is low, affecting the Martian climatic evolution. However, the behavior of the Martian climate system and the accompanying changes in climate and habitability of such planets remain ambiguous. Here we employed a latitudinally-resolved Martian energy balance model and assessed the possible climate on Mars for wider ranges of orbital parameters, solar irradiance, and total exchangeable $CO_2$ mass. We show that the atmospheric collapse occurs when the obliquity is below ~10˚ when other parameters are kept at the present Mars condition. We also show that the climate solutions on Mars depend on orbital parameters, solar luminosity, and the total exchangeable $CO_2$ mass. We found that the atmospheric collapse would have occurred repeatedly in the history of Mars following the variation of the axial obliquity, while the long-term evolution of atmospheric $p$CO₂ is also affected by the changes in the total exchangeable $CO_2$ mass in Martian history. Even considering the broad ranges of these parameters, the habitable conditions in the Martian $CO_2$-driven dry climate system would be limited to high-latitude summers.

## Highlights
– Atmospheric collapse occurs under obliquity lower than ~10˚ on the present Mars.
– Atmospheric collapses would have occurred repeatedly in the history of Mars.
– Habitability on Martian $CO_2$-driven dry climate systems would be limited.

**Introduction**

The climate of the present Mars is characterized by cold and dry conditions. Most of the surface $H_2O$ is located at the polar caps and the subsurface and $CO_2$ also condenses into the surface at high latitudes. There is no active volcanism and plate tectonics on the present Mars, meaning that the carbonate-silicate geochemical cycle, a negative feedback that stabilizes the atmospheric $pCO_2$ and climate in Earth's history (Walker et al. 1981), does not operate on present Mars. In such conditions, the Martian atmospheric $pCO_2$ and the climate are controlled by the equilibrium among the three exchangeable surface reservoirs of $CO_2$, that is, atmosphere, polar ice caps, and regolith (Gierasch and Toon 1973; Toon et al. 1980; McKay et al. 1991; Zent and Quinn 1995; Nakamura and Tajika 2003; Manning et al. 2006; Haberle et al. 2017; Jakosky and Edwards 2018; Buhler and Piqueux 2021). $CO_2$ condenses to the surface at high latitudes in winter and it sublimes into the atmosphere during the spring and the summer. Atmospheric $CO_2$ is also adsorbed by the surface regolith layer (Fanale and Cannon 1974; Toon et al. 1980; Fanale and Salvail 1982; Kahn 1985; Zent and Quinn 1995). The distribution of $CO_2$ among these surface reservoirs and climatic conditions are controlled by a thermodynamic balance between these surface reservoirs. This Martian $CO_2$-dominated climate system has two different stable steady states of climate modes under the present condition (Fanale and Salvail 1982; McKay et al. 1991; Haberle et al. 1994; Nakamura and Tajika 2001, 2002): one is the permanent-ice climate mode in which the atmospheric $pCO_2$ is controlled primarily by the thermodynamic balance between atmosphere and a permanent $CO_2$ ice and the other is the climate mode without a perennial $CO_2$ ice. It is noteworthy that, under the permanent-ice climate mode, the surface pressure is kept especially low because most $CO_2$ is partitioned into the $CO_2$ ice, hence the formation of permanent ice causes an event called "atmospheric collapse" (McKay et al. 1991; Haberle et al. 1994; Nakamura and Tajika 2002; Kreslavsky and Head 2005; Soto et al. 2015). Indeed, the time-dependent calculations of the Martian recent past indicate an occurrence of atmospheric collapse (Manning et al. 2006, 2019). Therefore, the behavior of the Martian $CO_2$ climate system is critical for understanding the evolution of the Martian surface environment.

One of the critical factors that affect the behavior of the Martian $CO_2$ climate system is the variations of orbital parameters (Toon et al. 1980; Nakamura and Tajika 2003; Kite et al. 2013, 2017; Soto et al. 2015; Batalha et al. 2016; Hayworth et al. 2020). It is well known that the variations of orbital parameters caused the quasi-periodic glacial-interglacial climate changes during the Quaternary on Earth (Hays et al. 1976; Raymo 1997; Huybers 2011; Abe-Ouchi et al. 2013; Watanabe et al. 2023). The variation of orbital parameters on Mars, especially that of obliquity, is characterized by its large amplitude of variations owing to the lack of a large moon (Laskar et al. 2004), which would have reached ~0° and ~60° in the past (Laskar et al. 2004) (Figure 1). The variation of obliquity modifies the latitudinal-seasonal distribution of the insolation received on the Martian surface (Toon et al. 1980; Nakamura and Tajika 2003; Soto et al. 2015). Specifically, the atmospheric collapse may have occurred when the obliquity was low in the recent past (Nakamura and Tajika 2003; Manning et al. 2006). This may have formed the layered deposits in the polar ice caps (Hvidberg et al. 2012; Bierson et al. 2016; Smith et al. 2016; Becerra et al. 2016, 2017, 2019; Lalich and Holt 2017; Manning et al. 2019; Lalich et al. 2019), surface deposits (Schon et al. 2009; Bernhardt et al. 2019), glaciers (Forget et al. 2006; Bramson et al. 2017; Dundas et al. 2018), and/or gullies at the mid-latitudes (Costard et al. 2002; Head et al. 2003; Conway et al. 2019).

In the more distant past, the solar irradiance and the total amount of exchangeable $CO_2$ would have been important in regulating the Martian climate. The solar luminosity was ~30% dimmer at ~4.5 Ga than the present brightness (Gough 1981). On the other hand, it has been suggested that there would have been an ocean on early Mars (Head et al. 1999; Citron et al. 2018), indicating a strong greenhouse effect for sustaining liquid water on the surface and dynamical climate evolution on early Mars (Batalha et al. 2016; Rapin et al. 2023; Kite and Conway 2024). The total exchangeable $CO_2$ must have been much higher on early Mars than on the present Mars. This would have helped warm the Martian surface by elevating the atmospheric $pCO_2$, while very high atmospheric $pCO_2$ would have promoted the formation of $CO_2$ clouds in the polar region (Kasting 1991; Pierrehumbert and Erlick 1998), potentially limiting the warming effect of $CO_2$. The total exchangeable $CO_2$ has decreased owing to the deposition of carbonates and escape to space via ion sputtering and photochemical escape (Kahn 1985; Haberle et al. 1994; Jakosky et al. 1994; Manning et al. 2006; Kurahashi-Nakamura and Tajika 2006; Hu et al. 2015). For a better constraint on the evolutionary path of the Martian climate, the behavior of the Martian $CO_2$ climate system to different values of orbital parameters, solar luminosity, and total mass of $CO_2$ is critical. However, the dependency of the climate mode for Mars and Mars-like exoplanets has been studied only for a very limited range of parameters (Nakamura and Tajika 2001, 2002, 2003; Soto et al. 2015). Here we employ a one-dimensional energy balance model and assess the multiple climate modes on Mars and Mars-like planets with different total exchangeable $CO_2$, solar irradiance, obliquity, precession, and eccentricity. We discuss possible Martian climate modes for wider ranges of parameters, considering Mars-like exoplanets which have a $CO_2$-dominated climate system similar to that of Mars, and discuss the habitability of such planets. We further show the condition of the atmospheric collapse on Mars and the time-dependent behavior of the atmospheric collapse.

**Method**

**One-dimensional energy balance model**

We used a latitudinally-resolved energy-balanced climate model (EBM) for Mars (Nakamura and Tajika 2003) (Figure 2). This model calculates the exchange of energy at the planetary surface and atmospheric layers and the exchange of $CO_2$ between the atmosphere, surface $CO_2$ ice, and surface regolith. The meridional resolution of the model is 2˚. In each grid, the atmosphere is divided into four layers that have an equal amount of atmosphere. $CO_2$ condensates onto the surface of each grid when the surface temperature is below the sublimation temperature. The balances of energy of the surface and atmospheric layers are represented as follows:

$$C_g \frac{dT}{dt} = (1 - \alpha) \cdot S + F_{ira,g} - F_{ire,g} - F_{sbl} + F_{lat} \qquad (1)$$

$$C_a \frac{dT_{a,1}}{dt} = F_{ira,a1} - F_{ire,a1} + F_{sbl} + F_{adv,a1} \qquad (2)$$

$$C_a \frac{dT_{a,i}}{dt} = F_{ira,ai} - F_{ire,ai} + F_{adv,ai} \qquad (i = 2, 3, 4) \qquad (3)$$

where the subscript $g$ and $a$ represent the ground layer and the atmospheric layer, respectively; the subscript number represents the number of the atmospheric layer (bottom layer is 1); $T$ is the temperature; $C_g$ and $C_a$ are the heat capacity of the ground and each atmospheric layer ($C_g = 1.0 \times 10^7$ J m$^{-2}$ K$^{-1}$), respectively (Nakamura and Tajika 2003); $S$ is the solar irradiance; $\alpha$ is the surface albedo; $F_{ire}$ and $F_{ira}$ are the emission and absorption flux of infrared radiation, respectively; $F_{sbl}$ and $F_{lat}$ are the sensible and latent heat flux, respectively.

In this model, we introduced the effect of the topographic dichotomy of Mars. The local surface pressure ($P_{atm,topo}$) is calculated as follows:

$$P_{atm,topo} = P_{atm} \cdot exp\left(-\frac{z}{H_1}\right) \qquad (4)$$

where $P_{atm}$ is the global-mean atmospheric $p\mathrm{CO_2}$, $z$ is the surface altitude relative to the mean topographic height of Mars, and $H_1$ is the local scale height of the atmospheric bottom layer, which is calculated assuming the hydrostatic equilibrium:

$$H_1 = \frac{kT_{a,1}}{mg} \qquad (5)$$

where $k$ is the Boltzmann constant, $m$ is the mean mass of the atmosphere (i.e., $CO_2$) ($m = 0.044$ kg mol$^{-1}$), and $g$ is the gravitational acceleration of Mars. The latitudinal mean topography of Mars is created using the dataset of the Mars Global Surveyor's Mars Orbiter Laser Altimeter (MOLA) (Zuber et al. 1992; Smith et al. 2001), MOLA Mission Experiment Gridded Data Records (MEGDRs).

The horizontal thermal advection terms in the atmospheric layer are represented as follows:

$$F_{adv,a,j} = \frac{D_j}{4(\Delta\Phi)^2 cos(lat_j)} \left(cos(\phi_{h,j+1}) \cdot (T_{a,j+1} - T_{a,j}) - cos(\phi_{h,j}) \cdot (T_{a,j} - T_{a,j-1})\right) \qquad (6)$$

Where $j$ is the latitudinal grid number ($j = 1, 2, \ldots$) and $\varphi$ is the latitude. The diffusion coefficient $D$ is calculated as follows:

$$D = D_0 \cdot P_{atm,topo}, \tag{7}$$

where $D_0$ is the diffusion coefficient at $P_{atm,topo}$ of 1 bar ($D_0 = 0.7$) (Stone 1972; Hoffert et al. 1981). The sensible heat flux is calculated as follows (Gierasch and Toon 1973; Yokohata et al. 2002; Nakamura and Tajika 2003):

$$F_{sbl} = k_{sbl,0} \cdot P_{atm,topo} \cdot \frac{T_g - T_{a1}}{T_{a1}} \tag{8}$$

where $F_{sbl,0}$ is the reference sensible heat flux ($k_{sbl,0} = 2.04 \times 10^3$ W m$^{-2}$ bar$^{-1}$) (Nakamura and Tajika 2003). The long-wave emission flux from the ground and each atmospheric layer are calculated as follows:

$$F_{ire,g} = \epsilon_g \sigma T_g{}^4 \tag{9}$$

$$F_{ire,ai} = 2\epsilon_a \sigma T_{a,i}{}^4 \qquad (i = 1, 2, 3, 4) \tag{10}$$

where $\sigma$ is the Stefan-Boltzmann constant ($5.67 \times 10^{-8}$ W m$^{-2}$ K$^{-4}$) and $\varepsilon_g$ is the ground emissivity ($\varepsilon_g = 1$). The atmospheric emissivity ($\varepsilon_a$) is parameterized to fit the surface temperature estimated by a previous study (Pollack et al. 1987), as follows:

$$\epsilon_a = \epsilon_{a1} y^5 + \epsilon_{a2} y^4 + \epsilon_{a3} y^3 + \epsilon_{a4} y^2 + \epsilon_{a5} y + \epsilon_{a6} \qquad (P_{atm,topo} \geq 0.0004 \text{ bar})$$
$$\tag{11}$$

$$\epsilon_a = \frac{P_{atm,topo}}{0.0004\,bar} \cdot \epsilon_a (P_{atm,topo} = 0.0004\,bar) \quad (P_{atm,topo} < 0.0004 \text{ bar})$$

where $y$ represents $\log(P_{atm,topo})$ and $\varepsilon_{a0}$, $\varepsilon_{a1}$, $\varepsilon_{a2}$, $\varepsilon_{a3}$, $\varepsilon_{a4}$, $\varepsilon_{a5}$, and $\varepsilon_{a6}$ are 0.0004, 0.009, 0.0792, 0.3324, 0.6782, and 0.5667, respectively. Using these values, $F_{ira}$ is represented as follows:

$$F_{ira,g} = \epsilon_g \cdot \epsilon_a \cdot (1 - \epsilon_a)^{j-1} \sigma T_{aj}{}^4 \tag{12}$$

$$F_{ira,ai} = F_{ira,ai\downarrow} + F_{ira,ai\uparrow} \tag{13}$$

The absorption fluxes of the downward and upward long-wave radiations in each atmospheric layer are represented as follows:

$$F_{ira,ai\downarrow} = \Sigma_{k=i+1,\ldots,4}[\epsilon_a(1 - \epsilon_a)^{k-i-1}\sigma T_{ak}{}^4] \qquad (i = 1, 2, 3) \tag{14}$$

$$F_{ira,ai\downarrow} = 0 \qquad\qquad (i = 4) \qquad (15)$$

$$F_{ira,ai\uparrow} = \epsilon_g \sigma T_g^{\ 4} \qquad\qquad (i = 1) \qquad (16)$$

$$F_{ira,ai\uparrow} = \epsilon_g (1 - \epsilon_a)^{i-1} \sigma T_g^{\ 4} + \Sigma_{k=1,\dots,i-1} [\epsilon_a (1 - \epsilon_a)^{i-k-1} \sigma T_{ak}^{\ 4}] \ (i = 2, 3, 4) \qquad (17)$$

where the subscripts $\downarrow$ and $\uparrow$ represent the absorption fluxes of downward and upward long-wave radiations, respectively.

The heat capacity of each atmospheric layer is calculated as follows:

$$C_a = 0.25 \cdot C_{a,0} \cdot P_{atm,topo} \qquad (18)$$

where $C_{a,0}$ is the heat capacity of the atmosphere at 1 bar ($C_{a,0} = 2.3 \times 10^7$ J m$^{-2}$ K$^{-1}$) (Nakamura and Tajika 2003). Note that the factor 0.25 is multiplied to divide the atmosphere into four layers that have an equal atmospheric mass. The surface albedos of the grid without and with $CO_2$ ice are defined here as $\alpha_g$ and $\alpha_i$, respectively, and represented as follows:

$$\alpha_g = \alpha_{g1}y^5 + \alpha_{g2}y^4 + \alpha_{g3}y^3 + \alpha_{g4}y^2 + \alpha_{g5}y + \alpha_{g6} \quad (P_{\text{atm,topo}} > 0.001 \text{ bar})$$

$$= \alpha_{g0} \qquad\qquad (P_{\text{atm,topo}} \leq 0.001 \text{ bar}) \quad (19)$$

$$\alpha_i = \alpha_{i1}y^3 + \alpha_{i2}y^2 + \alpha_{i3}y + \alpha_{i4} \qquad\qquad (P_{\text{atm,topo}} > 0.001 \text{ bar})$$

$$= \alpha_{i0} \qquad\qquad (P_{\text{atm,topo}} \leq 0.001 \text{ bar}) \quad (20)$$

where $\alpha_{g0}$, $\alpha_{g1}$, $\alpha_{g2}$, $\alpha_{g3}$, $\alpha_{g4}$, $\alpha_{g5}$, and $\alpha_{g6}$ are 0.21, –0.0008, –0.0074, –0.0147, 0.0337, 0.1381, and 0.3249, respectively, and $\alpha_{i0}$, $\alpha_{i1}$, $\alpha_{i2}$, $\alpha_{i3}$, and $\alpha_{i4}$ are 0.4, 0.0029, 0.0232, 0.0598, and 0.45, respectively. The $CO_2$-dependence of the surface albedo represents the effect of Rayleigh scattering when $P_{\text{atm,topo}}$ is high. This parameterization is based on Nakamura and Tajika (2003), but the values of $\alpha_{i0}$ and $\alpha_{i4}$ are lowered from the original values (0.7 and 0.75, respectively) to fit the timings of the appearance and disappearance of polar $CO_2$ ice caps. The updated values are consistent with those estimated from recent observations (Gary-Bicas et al. 2020). We note here that we do not consider the effect of $H_2O$ ice on the surface albedo because our focus is to assess the behavior of the Martian $CO_2$-driven climate system. This may lead to the overestimation of the surface temperature especially at the poles and underestimation of the size of the surface $CO_2$ ice reservoir when the obliquity is low: however, we expect that this would not strongly modify the behavior of $CO_2$ investigated in this study.

**Condensation of $CO_2$ to the surface**

The condensation of $CO_2$ ice to the surface occurs when the surface temperature drops below the sublimation temperature. Similarly, the sublimation of $CO_2$ ice from the surface occurs when the surface temperature exceeds the sublimation temperature. The model first estimates the energy balance without latent heat of the atmospheric $CO_2$ condensation ($F_{lat} = 0$). If the surface temperature drops below the sublimation temperature, atmospheric $CO_2$ condenses at the surface, and the amount of condensed $CO_2$ at the time step is calculated as follows:

$$\Delta P_{ice} = - \frac{[(1-\alpha)\cdot S + F_{ira,g} - F_{ire,g} - F_{sbl}] - C_g \cdot (T_{sub} - T'_g)}{L} \qquad (21)$$

where $L$ is the latent heat and $T'_g$ is the surface temperature calculated with a $F_{lat}$ of zero. The $CO_2$ sublimation temperature ($T_{sub}$) is represented as follows (Nakamura and Tajika 2003):

$$T_{sub} = T_{sub,4}y^4 + T_{sub,3}y^3 + T_{sub,2}y^2 + T_{sub,1}y + T_{sub,0}, \qquad (22)$$

where ($T_{sub,0}$, $T_{sub,1}$, $T_{sub,2}$, $T_{sub,3}$, $T_{sub,4}$) is (194.36, 26.451, 2.8593, 0.1814, 0.0046).

**Adsorption of $CO_2$ to surface regolith**

We estimated the size of the regolith reservoir using the formulation of Nakamura and Tajika (2001; 2002).

$$P_{reg} = \int_{-\pi/2}^{\pi/2} K_{reg} \ exp\left(-\frac{T}{T_d}\right) P_{atm,topo}{}^\gamma \cos(\phi)d\phi, \qquad (23)$$

where $K_{reg}$ is the tuning factor to reproduce the present regolith reservoir size ($K_{reg} = 72.8$ bar$^{-\gamma}$), $T_d$ is 35 K, and $\gamma$ is 0.275. The value of $K_{reg}$ was tuned so that the model reproduces the present atmospheric $pCO_2$ (6 mbar) under the present exchangeable $CO_2$ reservoir size (53.88 mbar) (Hu et al., 2015). Because the timescale for the equilibrium of the regolith reservoir would be much longer than the Martian year (Kieffer and Zent 1992), the behavior of the regolith reservoir is treated differently. The model calculates the annual-mean temperature and atmospheric $p$CO$_2$ of each grid by averaging the temperature and atmospheric $p$CO$_2$ of the grid in the past time steps corresponding to one Martian year, which is saved in the model at every 5 time steps. Finally, the total exchangeable CO$_2$ reservoir size ($P_{tot}$) is represented as follows:

$$P_{tot} = P_{atm} + P_{ice} + P_{reg}, \tag{24}$$

where $P_{atm}$ and $P_{ice}$ are the mass of CO$_2$ in the atmosphere and surface CO$_2$ ice, respectively.

**Definition of the habitability of the surface**

We examined the surface area fraction of habitable conditions, defined here as conditions for liquid water to be able to exist ($T > 273$ K) in a year and/or in the area on a planet. In order to quantify the habitability of Mars, we adopted the concept of net fractional habitability, a fraction of the surface area or the orbit of a planet that might be habitable (i.e., surface temperatures between 273 K and 373 K) (Spiegel et al. 2008):

$$f_{hab} = \frac{1}{t_{year}} \int_{-\pi/2}^{\pi/2} \{ \int_0^P H(\phi, t) \cdot dt \} \cos \phi d\phi, \tag{25}$$

where $t_{year}$ is the length of the year and $H$ is the habitability function. The habitability function is represented as follows:

$$H(\phi, t) = 1 \quad (273 \leq T(\phi, t) \leq 373)$$

$$= 0 \quad (T(\phi, t) < 273 \; or \; T(\phi, t) > 373). \tag{26}$$

We note here that we assumed that there is only a small amount of H$_2$O that does not affect the general behavior of the CO$_2$-driven dry climate system in the surface environment as in the present Mars.

**Experimental setup**

We conducted numerical experiments under various total exchangeable $CO_2$ (i.e., exchangeable amounts of $CO_2$ in the surface reservoirs), orbital parameters (obliquity, precession angle from the vernal equinox, and eccentricity), and solar luminosity and assessed the existence of the multiple equilibriums of Mars. We varied one of these parameters from the present Mars condition. We further conducted calculations with varying a set of obliquity and $P_{tot}$, under the solar luminosities of 1.00 and 0.75 times relative to the present value. We also varied a set of solar luminosity and $P_{tot}$ under the obliquity of $10°$. We further conducted a series of experiments assuming different ages. In this series of experiments, we changed the solar luminosity and the exchangeable $CO_2$ reservoir size. The solar luminosity is calculated as follows (Gough 1981; Feulner 2012) (Figure 3a):

$$S^* = \frac{5}{2 \cdot (\frac{Age_{Ga}}{4.57}) + 5} \qquad (27)$$

where $S^*$ is the solar luminosity relative to the present and $Age_{Ga}$ is the age in Ga ($Age_{Ga} = 0$ Ga at the present). The total exchangeable $CO_2$ reservoir size is estimated based on the previous study considering the escape of $CO_2$ to space and the deposition of $CO_2$ as carbonates (Hu et al. 2015) (Figure 3b).

In addition, we have conducted a set of calculations with idealized variations of the obliquity to elucidate the response of the atmosphere–ice cap system. For these experiments, we assumed artificial variations of Martian obliquity, which varies between 5 and $35°$ with a cosine curve, with different periods (10, 20, 30, 60, and 100 kyr). The values of other orbital parameters are fixed at the present Martian value. For this calculation, the regolith reservoir is not coupled to the model because our model cannot simulate the transient response of the regolith reservoir. The total exchangeable $CO_2$ reservoir is set to 0.1 bar in this experiment.

**Results**

**Reproduction of the present Martian condition**

The results for the present Mars condition are shown in Figure 4. The simulated seasonal variation of the surface temperature is broadly consistent with the simulations by three-dimensional general circulation models (Toon et al. 1980; Pollack et al. 1981, 1990, 1993; Wood and Paige 1992) (Figure 4b). The seasonal variations of the surface pressure near the landing site of Viking 1 and 2 show that the variations of the atmospheric $p$CO$_2$ from the winter to summer of the northern hemisphere are slightly underestimated (Figure 4d). This would be attributed to the treatment of albedo of the CO$_2$ ice: the observed values are different between hemispheres, while the model assumes an equivalent value for both hemispheres. In our model, the permanent CO$_2$ ice does not form in the present Martian condition. The present Martian condition is reproduced with a total exchangeable reservoir size of ~0.054 bar, most of which is distributed to the surface regolith layer. The present observation of the Martian poles indicates that there is virtually no CO$_2$ ice (Kieffer 1979; Byrne 2009; Titus and Cushing 2014; Piqueux et al. 2015; Haberle et al. 2017). Although it sometimes remains in summer, we expect that this would be associated with the local topographic structure, and the behaviors of the present Martian would be sufficient for representing the fundamental behavior of the Martian CO$_2$-driven climate system as shown in this study.

**Behaviors of the Martian climate system under different climate forcings**

We show the dependency of the steady states of the Martian climate system on different values of obliquity in Figure 5a–5c. When the obliquity changes from the present value (25.19˚), the global mean surface temperature does not strongly shift from the present value (200–210 K). When obliquity is below a critical value (~12˚), permanent ice forms in high-latitude regions. This boundary is broadly consistent with the estimate based on three-dimensional AGCM (Soto et al. 2015). When the obliquity is 12˚, the permanent ice exists only in the north pole, while the permanent ice forms in the south pole when obliquity is below 10˚ and atmospheric $p$CO$_2$ becomes small. With the permanent CO$_2$ ice in both hemispheres, the $CO_2$ ice cap reservoir gradually becomes the primary exchangeable $CO_2$ reservoir as the obliquity becomes small. This is because the size of the regolith reservoir decreases following the decline in the atmospheric $pCO_2$. This behavior is similar to the estimates by the other theoretical model (Buhler and Piqueux 2021). The minimum temperature of the planetary surface also becomes lower following the decline in the sublimation temperature owing to the lowering atmospheric $pCO_2$. When the obliquity is higher than the present value, the atmospheric $pCO_2$ does not strongly respond to the increase of obliquity, which is different from the result of the previous study that suggests a stronger sensitivity of atmospheric $pCO_2$ to the increase of obliquity (Buhler and Piqueux 2021).

The seasonal and latitudinal distributions of the surface temperature for different obliquities are shown in Figure 4b and 6. When the obliquity is lower than ~28° (Figure 5b), the highest surface temperature is achieved at low-latitude regions (red dots in Figure 5c and Figure 6a). In this condition, the maximum surface temperature is around 240 K (Figure 5b), which is below the freezing point of water. The maximum surface temperature increases owing to an increase in the obliquity when obliquity is over ~28° because the insolation at high-latitude regions becomes higher than in low-latitude regions, leading to the high summer temperature (Figure 6c). Under the obliquity higher than ~35°, some parts of high-latitude regions become warmer than the freezing point of water during summer (Figure 5b), meaning that this region is at least seasonally habitable. However, the habitable condition is limited to only high-latitude regions on the surface during summer.

The responses of the distribution of $CO_2$ and the surface temperature to different total exchangeable $CO_2$ ($P_{tot}$) under the present Martian orbital parameters are summarized in Figure 7a–7c. When $P_{tot}$ is below ~130 mbar, the regolith reservoir is the dominant $CO_2$ reservoir as in the case of the present Mars (Figure 7a). When $P_{tot}$ is over ~130 mbar, the atmosphere becomes the dominant $CO_2$ reservoir. The $CO_2$ ice cap reservoir was small compared with atmosphere and regolith reservoirs at any $P_{tot}$. The size of the regolith reservoir increases following the increase of $P_{tot}$ when $P_{tot}$ is below ~130 mbar, while it becomes less sensitive to $P_{tot}$ when $P_{tot}$ is over ~130 mbar, at which the regolith reservoir size is ~50 mbar. The atmospheric $CO_2$ reservoir, on the other hand, correlates positively to $P_{tot}$ at any $P_{tot}$ explored in Figure 7a. As a result, the global annual mean surface temperature increases when $P_{tot}$ increases and it reaches ~250 K when $P_{tot}$ is 1 bar (Figure 7b). The ice cap reservoir was smaller than regolith and atmospheric reservoirs at any $P_{tot}$, which reaches the highest value of ~1 mbar when $P_{tot}$ is ~100 mbar. The highest surface temperature was achieved at ~40˚S when $P_{tot}$ is between 20 and 600 mbar, while it is at the southern pole when $P_{tot}$ is out of this range.

The calculations with respect to different solar luminosity under the present Martian orbital parameters and total exchangeable $CO_2$ are shown in Figure 7d–7f. The atmospheric and ice cap $CO_2$ reservoirs increase following the increase of the solar luminosity, while the regolith reservoir decreases slightly. The regolith reservoir was ~40–50 mbar at any solar luminosity explored in Figure 7d. The ice cap reservoir was lower than ~0.6 mbar regardless of the solar luminosity. The atmospheric $pCO_2$ was ~1 mbar when the relative solar luminosity was 0.7 (corresponding to ~4.5 Ga condition) and it increased to ~10 mbar when the relative solar luminosity was 1.2. The global annual mean surface temperature increases with respect to the increase of solar luminosity. It is ~190 K when the relative solar luminosity is 0.7 and it increases to ~220 K when the relative solar luminosity is 1.2. The solar luminosity did not strongly affect the latitude of the highest surface temperature.

The calculations with respect to different total exchangeable $CO_2$ and obliquity under the present solar luminosity are shown in Figure 8a. In this figure, the maximum $P_{atm}$ that is possible in each condition is shown with colors. As investigated above, the changes in $P_{atm}$ were controlled primarily by changes in $P_{tot}$ when the permanent $CO_2$ ice does not exist, while obliquity does not strongly affect $P_{atm}$. When $P_{tot}$ is sufficiently high, a climate mode without a permanent ice sheet can exist owing to the high atmospheric $pCO_2$ even under low obliquity conditions. Specifically, if obliquity is lower than ~12˚, the model expects that multiple climate modes can exist depending on the presence or absence of the permanent ice. Under the permanent ice solution, the atmospheric $pCO_2$ is similar to the values in the seasonal ice solution near the boundary of the multiple solutions.

In this case, the permanent ice exists only in the northern hemisphere, indicating that the formation of the southern $CO_2$ ice cap is important for the atmospheric collapse. The $f_{hab}$ value is shown in Figure 8d. Even with high $P_{tot}$ (1 bar) and high obliquity, the $f_{hab}$ value is smaller than 15%, indicating the limited surface habitability in the Mars-like $CO_2$ system. When the obliquity is below 20˚, $f_{hab}$ is zero even when $P_{tot}$ is 1 bar. Specifically, there is no latitudinal zone that is habitable throughout the year, and there is no season in which any latitude is habitable in our parameter space (Figure 8). These results indicate that the habitability on the surface of Mars would be limited. A larger atmospheric greenhouse effect than a 1-bar $CO_2$ atmosphere is necessary to keep the surface habitable as previously suggested (Pollack et al. 1987; Ramirez et al. 2014, 2020; Wordsworth et al. 2017; Kamada et al. 2020, 2021, 2022).

The responses of the Martian $CO_2$-driven system when eccentricity and precession angle from the vernal equinox are changed from the present Martian conditions are shown in Figures 5d–5f and 5g–5i, respectively. The distributions of the $CO_2$ reservoirs do not respond to the change in eccentricity, even under unrealistically high eccentricity (~0.4). The latitude with the highest surface temperature was ~40˚N when eccentricity is zero, while it turns to ~40˚S when eccentricity is over zero (Figure 5f). Higher eccentricity increases the maximum temperature, while the global mean surface temperature does not respond to eccentricity. When the precession angle from the vernal equinox is changed, the distributions of the $CO_2$ reservoirs do not respond to the change in eccentricity (Figure 5g). The latitude with the highest surface temperature, on the other hand, depends on the precession angle. This indicates that obliquity would have a dominant control in the atmospheric $pCO_2$ than eccentricity and precession. When the precession angle from the vernal equinox is ~10–180˚, it is ~30–40˚N, while it turns to ~30–40˚S when the precession angle from the vernal equinox is ~180–360˚.

We further show the results under different values of solar irradiance and total $CO_2$ mass on the climate system (Figure 8c). Calculations were conducted with respect to low obliquity value (10˚). With the lower $P_{tot}$ and solar luminosity, the atmospheric $pCO_2$ becomes low owing to the accumulation of $CO_2$ ice. When $P_{tot}$ is sufficiently high, the climate mode without permanent $CO_2$ ice can exist owing to the stronger greenhouse effect. This boundary is also dependent on the solar luminosity. With high solar luminosity, climate mode without permanent ice can be achieved with lower $P_{tot}$. The $f_{hab}$ value is zero even with $P_{tot}$ of 1 bar when the solar luminosity is below the present value, supporting the difficulty in sustaining the warm environment on early Mars only with $CO_2$.

**Evolution of the Martian $CO_2$ system during the Noachian and Hesperian**

The result of calculations with respect to different ages considering the changes in $P_{tot}$ and solar luminosity in the history of Mars, the distribution of $CO_2$ is estimated in Figure 9 for four different obliquity values (10, 25.19, 45, and 60˚). For the case with the present obliquity, the initial atmospheric $pCO_2$ at 3.8 Ga is ~0.2–0.3 bar, which was the largest reservoir of $CO_2$ (Figures 9a and 9b). The atmospheric $pCO_2$ decreases following the decline in $P_{tot}$ caused by ion sputtering and photochemical escape and deposition of carbonates, while the changes in the $CO_2$ ice and regolith reservoirs do not respond strongly to the decline in $P_{tot}$. The decline in atmospheric $pCO_2$ ceases at ~3.0 Ga when the $P_{tot}$ becomes close to the present value. Since then, the distribution of $CO_2$ has not varied strongly, indicating that the $CO_2$ system is less sensitive to the changes in the solar luminosity in the history of Mars. The overall characteristics of the responses are similar for the case with a higher obliquity value (45˚ and 60˚). For the case of low obliquity (10˚), the atmospheric $pCO_2$ was ~1 mbar at 3.8 Ga, resulting in a lower surface temperature. In this case, the mass of $CO_2$ ice gradually decreases until 1.0 Ga owing to the decline in $P_{tot}$ and the increase in solar luminosity. The atmospheric $pCO_2$ drops down to ~0.3 mbar at ~3.0 Ga and it gradually increases following the increase in the solar luminosity. These results indicate that the values of obliquity are especially important in understanding the early history of Mars. Specifically, the atmospheric $pCO_2$ may fluctuate between >100 and <1 mbar on early Mars owing to the variation of obliquity. This means that the atmospheric collapse that results in the change in atmospheric $pCO_2$ with a factor of more than 100 could have occurred repeatedly following the obliquity cycle.

**Conditions and timescales of the atmospheric collapse**

The conditions for the atmospheric collapse under two different solar luminosities (75 and 100% of the present brightness) are shown in Figures 8b and 8a, respectively. When the solar luminosity is 75% (Figure 8b), the atmospheric collapse occurs only when the total exchangeable $CO_2$ becomes less than 0.3–0.6 bar when the initial state is set to the condition with high atmospheric $pCO_2$. With a larger total exchangeable $CO_2$, the atmospheric collapse cannot occur owing to the greenhouse effect of atmospheric $CO_2$. The atmospheric collapse also occurs when the obliquity becomes low. The critical obliquity increases up to ~20˚ following the increase of $P_{tot}$ under a total exchangeable $CO_2$ below ~0.1 bar (Figure 8b). With a solar luminosity of 100 % (Figure 8a), the maximum values of both total exchangeable $CO_2$ and obliquity for the atmospheric collapse are smaller than for the case with 75 % solar luminosity. These results indicate that atmospheric collapse occurs more easily on early Mars than on present Mars. The Martian past obliquity would have transiently dropped below 22˚ in the Martian recent past (Laskar et al. 2004) (Figure 1a), which would have caused the atmospheric collapse on early Mars.

The time-dependent behavior of the atmosphere-ice cap system when the model is run under a transient variation of obliquity and the present solar luminosity is shown in Figure 10. When the obliquity drops below the critical value for the atmospheric collapse (~12˚), the atmospheric $pCO_2$ suddenly decreases from the initial value (0.1 bar) to levels lower than $10^{-4}$ bar. After the collapse of the atmosphere, the atmosphere eventually starts to increase following the increase in the obliquity. The atmospheric $pCO_2$ returns to the initial value ~20 kyr after the timing of the lowest atmospheric $pCO_2$. These results show that the response timescale of the atmospheric collapse is well shorter than the variations of orbital parameters. This indicates that the atmospheric collapse would occur repeatedly given that the obliquity drops repeatedly below the critical level for the atmospheric collapse.

**Discussion**

In this study, we investigated the climate and habitability of Mars under dry conditions. We showed that once the Martian obliquity decreases below the threshold obliquity for the formation of the southern permanent ice, the atmospheric $p\mathrm{CO_2}$ would decrease sharply, which corresponds to the atmospheric collapse (Soto et al. 2015). This boundary obliquity is lower than the reconstructed obliquity for the last 10 Myr, indicating that the atmospheric collapse would not have occurred in the last 10 Myr (Figure 1). A previous modeling study indicated that the layered $\mathrm{CO_2}$ ice in the southern ice cap recorded periods of low obliquity in the recent past (Buhler et al. 2020; Buhler 2023). On the other hand, our model infers that the permanent $\mathrm{CO_2}$ ice did not form in the recent past. This may suggest that the layered $\mathrm{CO_2}$ ice formed regionally owing to the local topographic reliefs and developments of stationary waves or it may have formed in a more distant past. For a better constraint on the formation of the perennial $\mathrm{CO_2}$ ice, investigations using three-dimensional models would be desirable (Soto et al. 2015). In the more distant past, variations of Martian obliquity are difficult to estimate owing to the chaotic behavior of orbital parameters, however, results of the ensemble of the estimates of the obliquity for the last 250 Myr exhibit periods with obliquity lower than 10° (Laskar et al., 2004). In such conditions, the atmospheric collapse would have occurred repeatedly on Mars.

Our result indicates that the atmospheric $p\mathrm{CO_2}$ would have decreased following the decline in the total reservoir size of exchangeable $\mathrm{CO_2}$ via ion sputtering and photochemical escape to space and/or deposition of carbonate. Therefore, the long-term history of the outgassing of $\mathrm{CO_2}$, escape of $\mathrm{CO_2}$, and the deposition of carbonate would be central to understanding Martian $\mathrm{CO_2}$ distributions among the atmosphere, regolith, and ice caps. The carbonate deposition during the Noachian and Hesperian is highly uncertain. The estimate based on the exposure at Nili Fossae is 0.012 bar (Ehlmann et al. 2008; Edwards and Ehlmann 2015; Hu et al. 2015). The upper estimate based on the non-detectability of carbonate from orbital remote sensing data is 1.4 bar, assuming the 500 m thickness of carbonate-bearing crust, while this value would be down to 0.3 bar with a plausible carbonate content in the crust (Hu et al. 2015). The deposition of carbonate during the Amazonian is estimated to be at most 7 mbar, based on the carbonate content in Martian dust and soil of the northern plain (Bandfield et al. 2003; Zuber et al. 2007; Watters et al. 2007; Sutter et al. 2012; Leshin et al. 2013; Hu et al. 2015). Similarly, the $CO_2$ outgassing flux on Martian history is affected by the outgassing scenarios and other factors such as the crustal redox state (Tajika and Sasaki 1996; Craddock and Greeley 2009; Grott et al. 2011). For a better constraint on the history of the Martian surface environment, further investigations of these processes together with the consideration of carbon isotopes would be beneficial for further investigation (Hu et al. 2015).

We showed that the surface habitability on Mars during the Amazonian is very limited. For such planets, life, if existed, may be suited for inhabiting subsurface environments because terrestrial microbes are not able to control their body temperature (Wright and Cooper 1981; Huey and Kingsolver 1989; Barnett and Olson 2022). Nevertheless, our results indicate that the surface becomes habitable at least seasonally with a wide range of total atmospheric $p$$CO_2$ when obliquity and/or eccentricity is high. The potential biospheric-atmospheric interactions with subsurface ecosystems would be a fruitful topic for future research. It is noteworthy that the actual habitability on Mars depends not only on climate and availability of water on the planetary surface but also on other factors such as supplies of essential elements, atmospheric redox state, and/or conditions suitable for the production of organic matter (Cockell 2014; Ehlmann et al. 2016; Koyama et al. 2021; Watanabe and Ozaki 2024; Ueno et al. 2024). Investigating the combination of these processes together with the advancement of future Mars missions for life detection (Jones 2018; Changela et al. 2021) would further help in understanding the habitability of Mars-like planets for future searches for signs of life on exoplanetary atmospheres.

## Conclusions

The stability and habitability of the Martian $CO_2$-driven climate system were investigated in this study. We conducted numerical simulations under various conditions of orbital parameters, solar irradiance, and the total mass of surface exchangeable $CO_2$ using an energy-balanced climate model. The climate solutions on Martian $CO_2$-driven climate systems depend on orbital parameters, solar luminosity, and the total $CO_2$ mass. Even considering the broad ranges of these parameters, the habitable condition on the Martian surface would be limited to high-latitude summer. The atmospheric collapse occurs when the obliquity is low, indicating that it would have occurred repeatedly in the history of Mars.

**Acknowledgments**

We thank H. Kurokawa for fruitful discussion. We also thank S. Kadoya for supporting the construction of the model. AK is supported by the Fusion Oriented REsearch for disruptive Science and Technology (FOREST) Program of the Japan Science and Technology Agency (JST) (Grant Number JPMJFR212U), the Japan Society for the Promotion of Science (JSPS) KAKENHI Grant Numbers JP23K13166.

**Authorship statement**

Conceptualization: YW; Methodology: YW; Software: YW; Investigation: YW, AK; Funding Acquisition: AK; Supervision: YW, ET; Visualization: YW; Writing – original draft: YW; Writing – review & editing: YW, ET, AK

**Conflict of Interest Statement**

The authors declare that they have no competing interest.

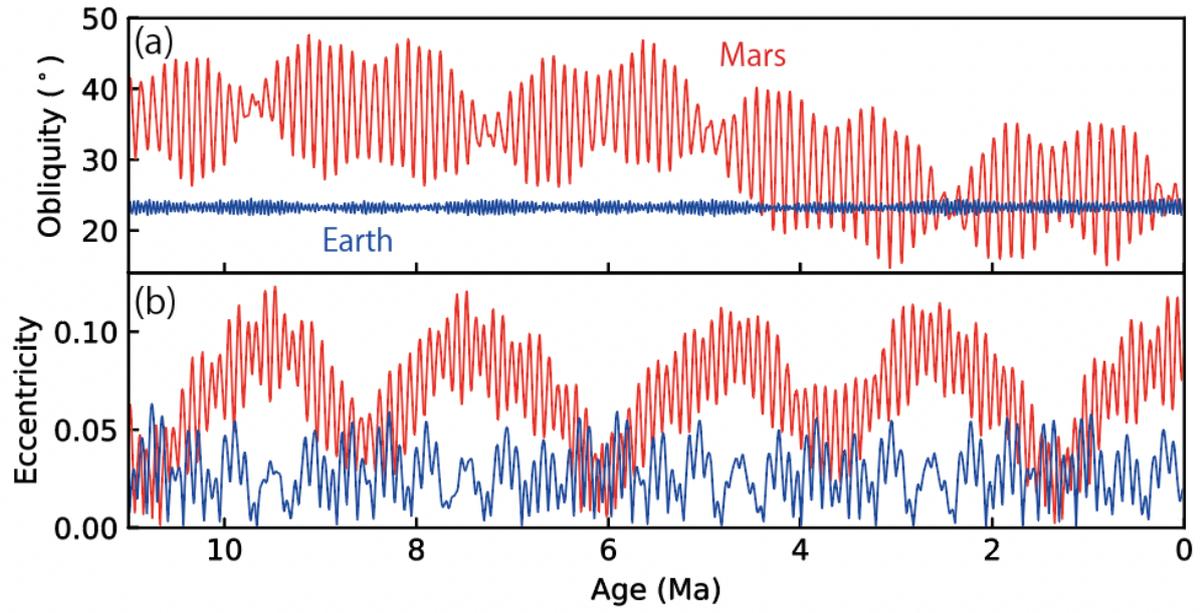

**Figure 1.** Variations of obliquity (a) and eccentricity (b) of Earth and Mars in the recent past (blue and red lines, respectively).

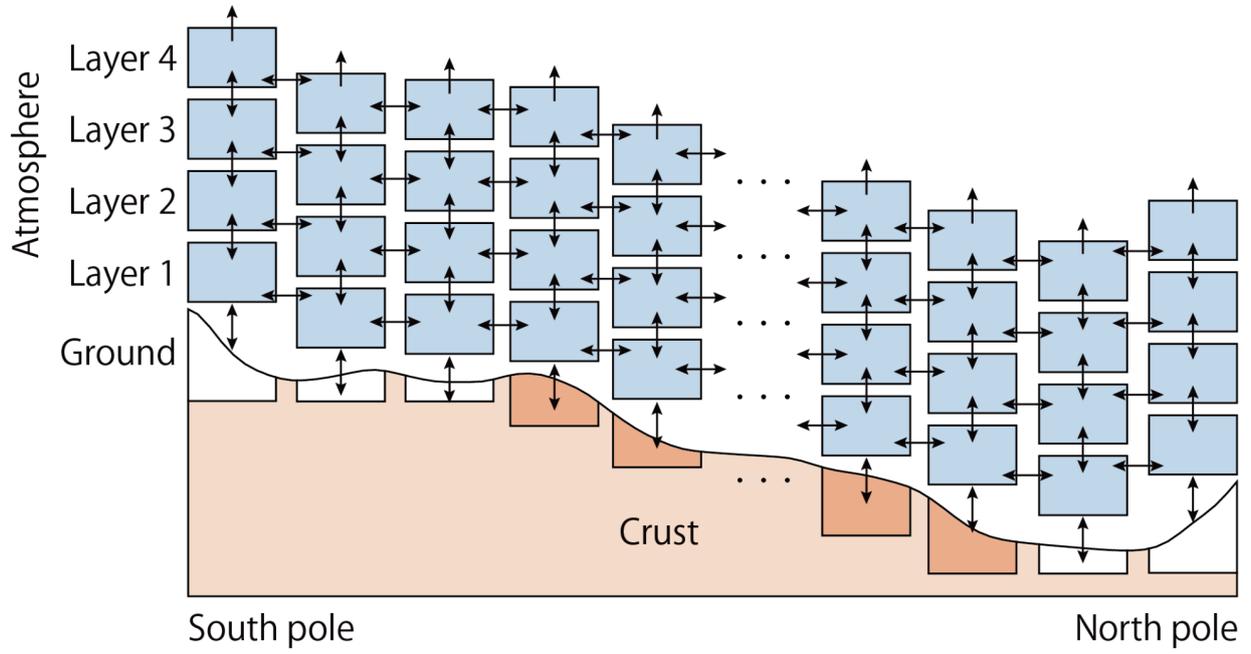

**Figure 2.** Schematic illustration of the energy-balanced climate model used in this study.

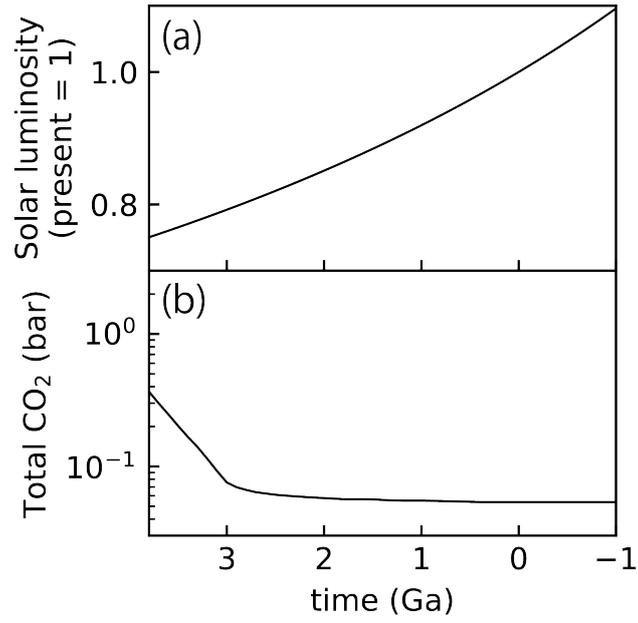

**Figure 3.** Evolutions of the solar luminosity (Gough, 1981) (a) and the total exchangeable $CO_2$ reservoir (atmosphere, ice, and regolith) (Hu et al., 2015) (b).

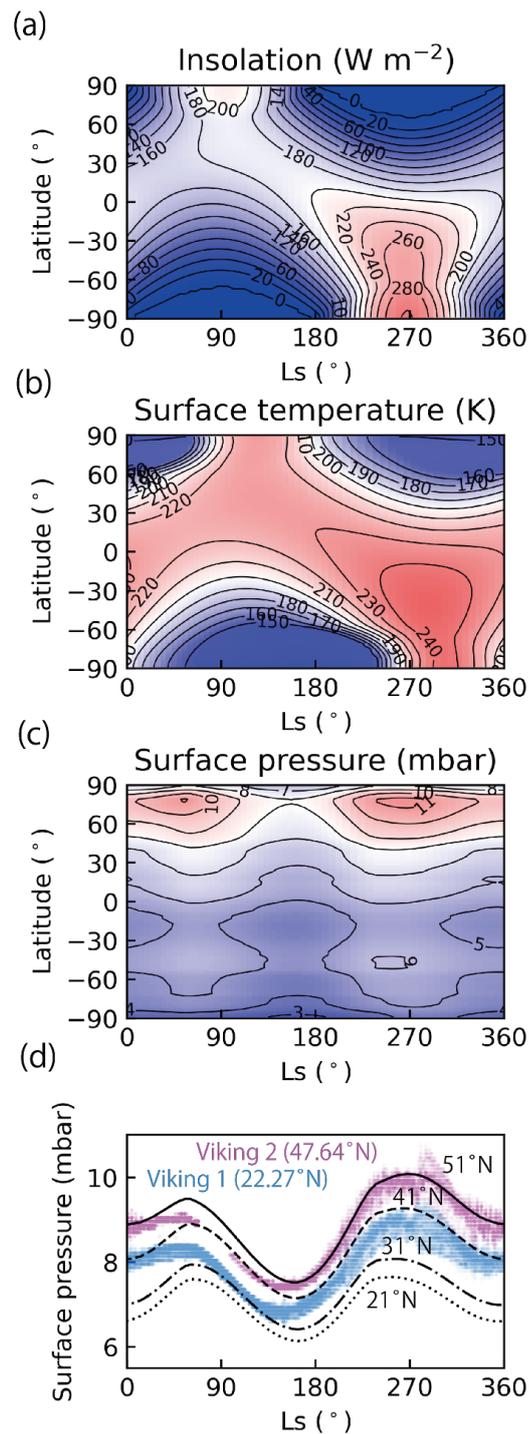

**Figure 4.** Seasonal and latitudinal variations of insolation (a), surface temperature (b), and surface pressure (c) simulated in our model assuming the present Mars condition. (d) Variations of the surface pressure simulated in our model assuming the present Mars condition at latitudes near the landing sites of Viking 1 and 2. Blue and purple represent observed variations of the surface pressure at landing sites) (Tillman, 1989).

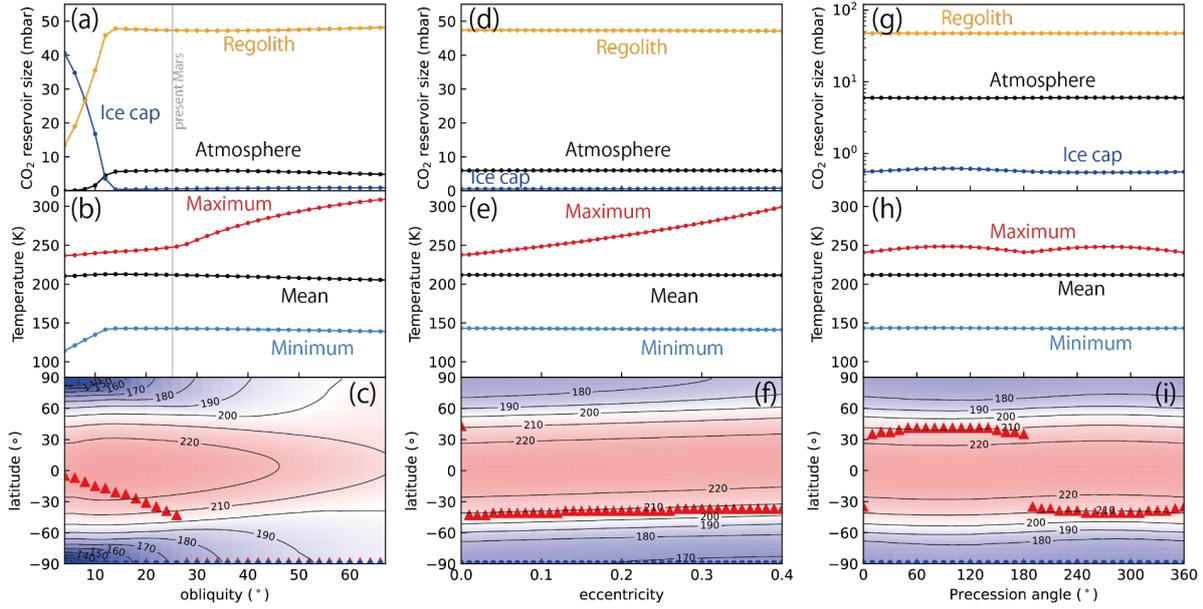

**Figure 5.** Distributions of $CO_2$ between the atmosphere, ice, and regolith (black, blue, and orange lines) (a, d, g), the maximum, mean, and minimum temperatures that are achieved on a planet (red, black, and blue lines, respectively) (b, e, h), and the latitudinal mean temperature (background color) and latitudes of maximum and minimum temperatures (red triangles and blue dots, respectively) (c, f, i) with respect to different obliquity (a–c), eccentricity (d–f), and precession angle from the vernal equinox (g–i).

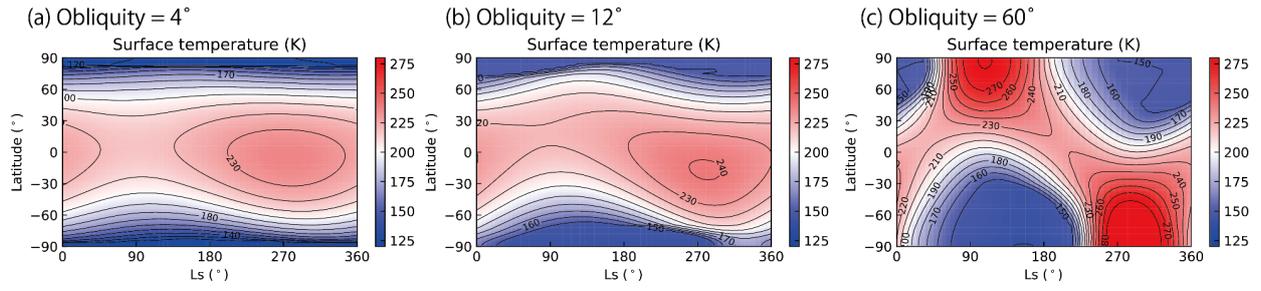

**Figure 6.** Seasonal and latitudinal variations of the surface temperature simulated with different obliquity (4, 12, and 60˚ for a, b, and c, respectively).

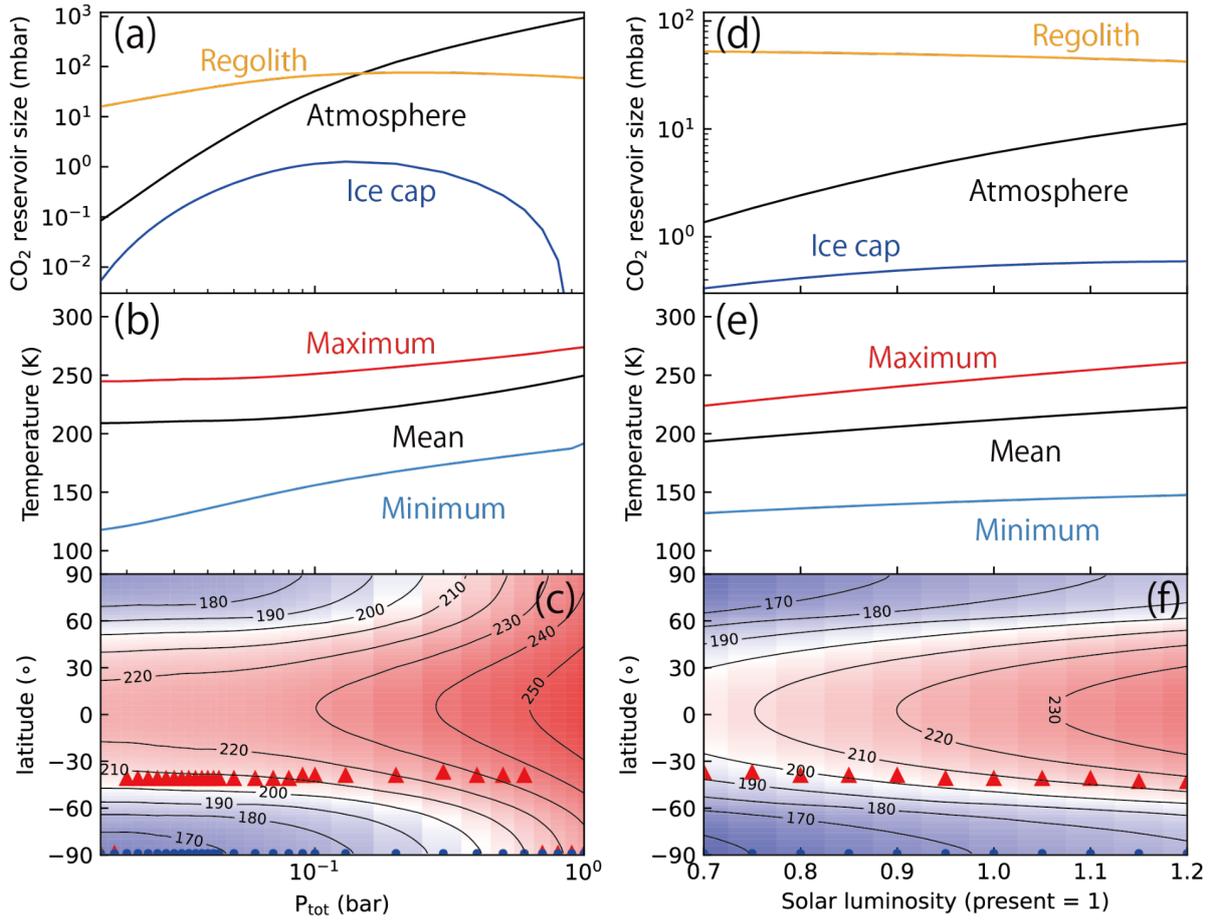

**Figure 7.** Distributions of $CO_2$ between the atmosphere, ice, and regolith (black, blue, and orange lines) (a, d), the maximum, mean, and minimum temperatures that are achieved on a planet (red, black, and blue lines, respectively) (b, e), and the latitudinal mean temperature (background color) and latitudes of maximum and minimum temperatures (red triangles and blue dots, respectively) (c, f) with respect to different total exchangeable $CO_2$ ($P_{tot}$) (a, b) and different solar luminosity (present = 1) (c, d). Calculations are conducted with the present Martian orbital parameters.

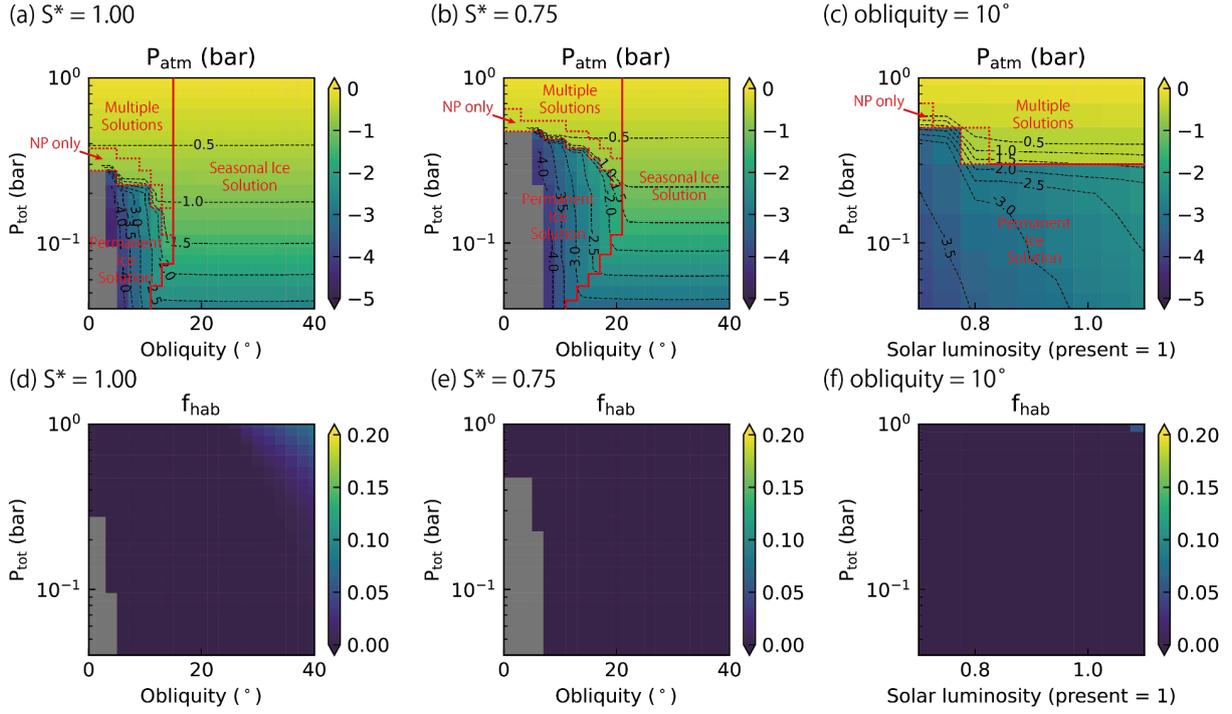

**Figure 8.** $CO_2$ reservoir sizes of the atmosphere ($P_{atm}$) (a–c), and the fractional habitability ($f_{hab}$) (d–f) calculated with respect to different obliquity and total exchangeable reservoir size ($P_{tot}$) under present solar luminosity (a, d), different solar luminosity and $P_{tot}$ under an obliquity of 10° (b, e), and different obliquity and $P_{tot}$ under solar luminosity of 0.75 times relative to the present value (c, f). The region "NP only" represents the solution with the permanent ice only at the north pole. The $P_{atm}$ values shown in a, b, and c are the maximum value that is possible in each condition.

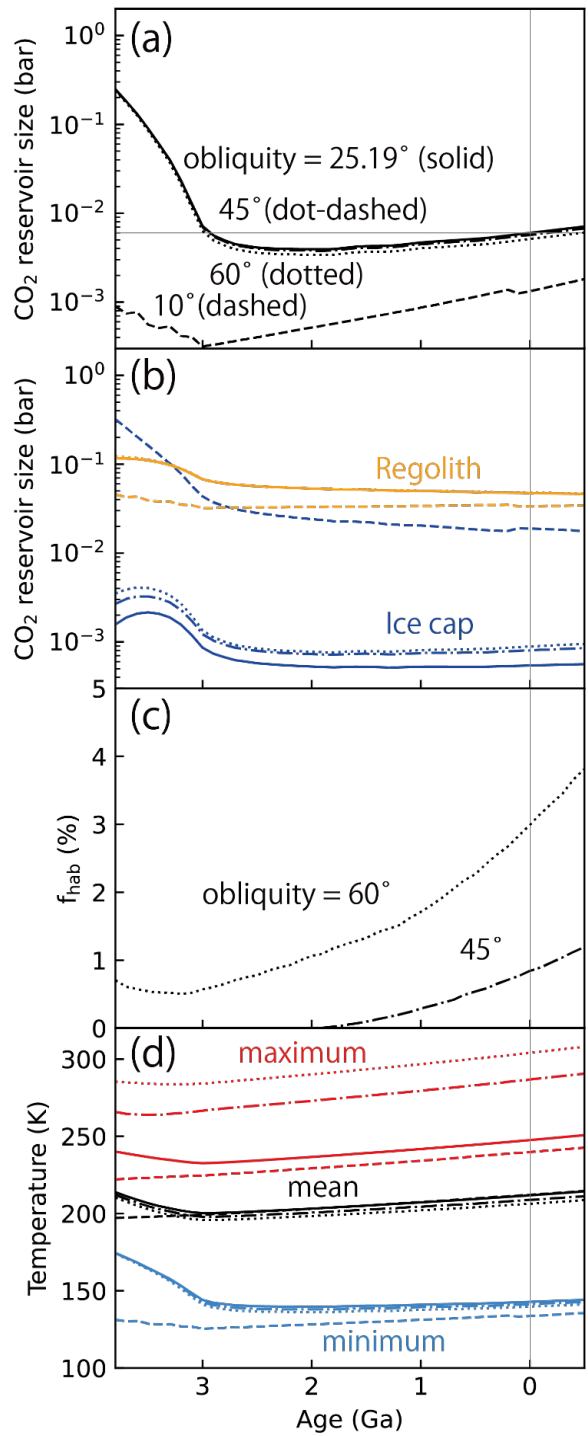

**Figure 9.** Evolutions of the exchangeable reservoirs and climate of Mars. (a) Atmospheric $p$CO$_2$, (b) ice and regolith reservoirs, (c) fractional habitability, and (d) maximum, mean, and minimum temperatures that are achieved on Mars. Calculations were conducted with respect to four obliquity values (10, 25.19, 45, and 60˚; dashed, solid, dot-dashed, and dotted lines, respectively).

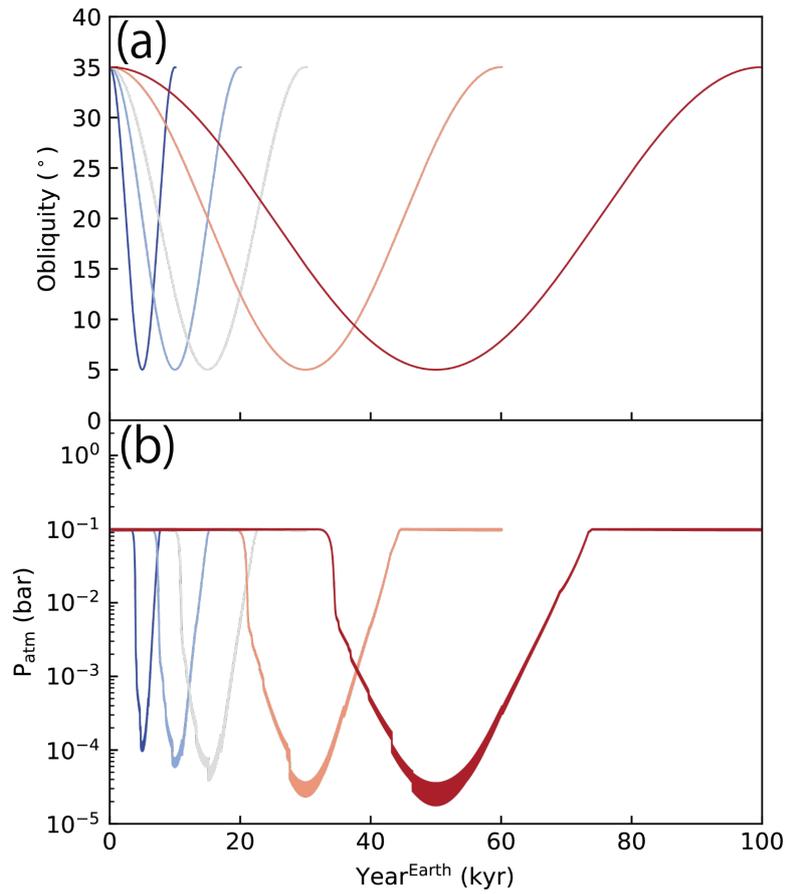

**Figure 10.** Time-dependent response of the Mars-like atmosphere-ice cap system to variations of obliquity forcing with different periodicities. (a) Variations of the obliquity forcing and (b) and corresponding variations of the Mars-like atmosphere-ice cap system are shown with solid lines. Colors represent the different periodicity of the variations of obliquity assumed in each calculation.